\documentclass[apj]{emulateapj}

\usepackage{hyperref}

\begin{document}

\title{Magnetic Structure of Rapidly Rotating FK Comae-Type Coronae}

\author{O. Cohen\altaffilmark{1}, J.J. Drake\altaffilmark{1}, V.L. Kashyap\altaffilmark{1},
H. Korhonen\altaffilmark{2},D. Elstner\altaffilmark{3}, T.I. Gombosi\altaffilmark{4}}

\altaffiltext{1}{Harvard-Smithsonian Center for Astrophysics, 60 Garden St. Cambridge, MA 02138}
\altaffiltext{2}{European Southern Observatory, Karl-Schwarzschild-Str. 2, D-85748 Garching bei
Muenchen,  Germany}
\altaffiltext{3}{Astrophysikalisches Institut Potsdam, An der Sternwarte 16, D-14482 Potsdam, Germany}
\altaffiltext{4}{Center for Space Environment Modeling, University of Michigan, 2455 Hayward St.,
Ann Arbor, MI 48109}

\begin{abstract}

We present a three-dimensional simulation of the corona of an FK~Com-type rapidly rotating G giant using a
magnetohydrodynamic model that was originally developed for the solar corona in order to
capture the more realistic, non-potential coronal structure. We drive the simulation with
surface maps for the radial magnetic field obtained from a stellar dynamo model of
the FK~Com system.  This enables us to obtain the coronal structure for different field topologies
representing different periods of time.  We find that the corona
of such an FK~Com-like star, including the large scale coronal loops, is dominated by a strong toroidal component of the magnetic field.  This is a result of 
part of the field being dragged by the radial outflow, while the other part remains attached to the rapidly 
rotating stellar surface.  This tangling of the magnetic field,
in addition to a reduction in the radial flow component, leads to a 
flattening of the gas density profile with distance in the inner part of the corona. The three-dimensional simulation provides a
global view of the coronal structure.  Some aspects of the results, such as the toroidal wrapping of the magnetic field, should also be 
applicable to coronae on fast rotators in general, which our study shows can be considerably different from the well-studied
and well-observed solar corona.  Studying the global structure of such coronae should
also lead to a better understanding of their related stellar processes, such as flares and coronal mass ejections, and in
particular, should lead to an improved understanding of mass and angular momentum loss 
from such systems.

\end{abstract}



\section{INTRODUCTION}
\label{sec:Intro}

A massive object that rotates very quickly and whose magnetic field flip-flops can significantly affect
the surrounding environment. Such an object is the single, late-type giant G star
FK~Comae~Berenices (FK~Com; HD\,117555), which is the eponymous prototype for the so-called ``FK~Com-type stars''
\citep{BoppStencel81}.  The FK~Com stars are observed to have extremely fast rotation
accompanied by enhanced stellar activity and are thought to have evolved from coalesced binaries.

FK~Com itself has been observed since the 1960s, and at the beginning of the 1990s
\cite{Jetsu91,Jetsu93} presented a long-term study of the magnetic activity of the system.
They found that over a period of 25 years the location of the most concentrated 
spot activity has been moving back and forth between the same two longitudes, which are
separated by 180 degrees. They nicknamed this activity behavior a ``Flip-Flop'' of the
stellar magnetic field. 

In the past two decades, many observations of the FK~Com system have been undertaken to better
resolve the stellar flip-flop phenomenon.  In a series of papers,
\cite{KorhonenI,KorhonenII,KorhonenIII,KorhonenIV,KorhonenV} and \cite{KorhonenVI} have performed
spectroscopic and photometric studies of the evolution of surface spot activity on FK~Com, while \cite{KorhonenElstner05}
studied this evolution using a stellar dynamo model. Observations across the electromagnetic spectrum, including radio bands \citep[e.g][]{Hughes87,Rucinski91}, reveal 
broad $H\alpha$ emission
\citep[e.g][]{Ramsey81,Walter82,Holtzman84,Kjurkchieva05}, strong UV emission in which transition region lines are broadened to full-width 
half-maxima in excess of $500\;km\;s^{-1}$, which is at least twice the projected surface equatorial rotation velocity \citep{Ayres06}, 
and X-ray activity greatly elevated compared with normal stars at this age and stage of evolution \citep[e.g][]{Walter81,Drake08,Buzasi03}

The rapid rotation of FK~Com is probably the source of the observed enhanced activity, while the
longitudinal flip-flop of the active spot area is related to stellar dynamo action. The flip-flop
does not necessarily correspond to a complete reversal of the magnetic field, but rather a
reversal in the location of particular, more active longitudes due to a non-axisymmetric stellar dynamo \citep[e.g][]{Berdyugina04}.  
Evidence for magnetic flip-flop behavior on the Sun \citep{Berdyugina03}
suggests that this phenomenon might be a universal feature of any realistic, non-axisymmetric stellar dynamo.

The magnetic flip-flop by itself is intriguing.  Here, however, we ask what would be the coronal 
structure of such fast rotating, flip-flopping star? In
the case of slowly rotating stars like the Sun, the structure of the stellar corona and the
stellar wind is dominated by the topology of the stellar magnetic field. A common approximation
for describing both solar and stellar coronae is the so-called ``potential field approximation''.
Under this approximation, there is no forcing on the magnetic field (i.e., magnetostatic field)
so the magnetic field can be described as a gradient of a scalar potential, and the three
dimensional distribution of the magnetic field can be obtained by solving Laplace's equation
for this scalar potential.

This solution requires two boundary conditions for the spatial
domain.  The inner boundary condition can be obtained using available surface maps of the stellar
magnetic field. The outer boundary condition, however, cannot be obtained from observations.  It is
set in a more arbitrary manner, by the requirement that at this boundary the magnetic field is purely radial. This assumption is reasonable if we consider
that the stellar wind overcomes the magnetic pressure of the stellar magnetic field, so the latter
becomes fully open above the Alfv\'en point (at which the Alfv\'enic Mach number, $M_A=u/u_A=1$,
with $u_A=B/\sqrt{4\pi\rho}$ being the Alfv\'en speed). It is convenient to define a surface
(``source'' or Alfv\'enic surface), which is the manifold defined by the Alfv\'enic points. 
This is commonly approximated using a spherical surface with a certain height above the stellar
surface, although the correctness of such an approximation has been under debate even for the solar case \citep[e.g][]{Riley06,Gilbert07}.
An even more serious problem arises when the very basic assumption, that the forces in the system
are negligible compared to the magnetic force, breaks down.  The rotation period of FK~Com
is 2.4 days \citep{BoppRucinski81,WalterBasri82,KorhonenII,Ayres06}, and the equatorial azimuthal speed, $u_\phi=r\Omega_\star$, in the low corona ($1-3.5R_\star$)
is about $250-700\;km\;s^{-1}$.  At these speeds, the dynamic pressure is 
$p_{dyn}=\rho u^2_\phi \approx 0.01-1\;dyne\;cm^{-3}$. Even if we consider a strong field on the surface
with a field strength of $B_0=250\;G$, the magnetic pressure at $r=3.5R_\star$ will be of the order of
$P_m=B^2_0/(3.5^6\cdot 8\pi)\approx 1\;dyne\;cm^{-3} $. Therefore, in these rapidly rotating systems,
the ratio of $p_{dyn}/p_m$ is not very small, and the potential field approximation is no longer valid.  A more physical method to 
describe the coronal structure is required.

In this paper, we describe a magnetohydrodynamic (MHD) simulation of the corona of an FK~Com-like star based on the flip-flop dynamo model of 
the surface magnetic field of \citet{Elstner05} and \citet{KorhonenElstner05}.
We adapt a global MHD model, originally developed for the solar corona, to the FK~Com model system.
This model enables us to study the global structure of the corona and to obtain the steady-state,
non-potential solution that includes the stellar wind and the effects of rapid rotation.
Another advantage of the MHD solution over the potential field approximation is that it provides
the solution for the complete set of physical parameters describing the coronal plasma: the magnetic field, gas density, temperature, velocity and electric currents throughout the three-dimensional volume under consideration, while the potential field only provides 
information about the magnetic field distribution.

The paper is structured as follows.  The numerical model and
the observational constraints used in the simulation are described in Section~\ref{sec:Simulation}.
The results are presented in Section~\ref{sec:Results}, and the main findings are discussed 
in Section~\ref{sec:Discussion}. We summarize this work in Section~\ref{sec:Conclusions}.\\\\
\newline


\section{NUMERICAL SIMULATION}
\label{sec:Simulation}

\subsection{MODEL DESCRIPTION AND OBSERVATIONAL CONSTRAINTS}
\label{sec:Model1}

For the simulation of an FK~Com-like star, we use the solar corona model by \cite{cohen07, cohen08b},
which is part of the Space Weather Modeling Framework (SWMF) \citep{toth05} and is based
on the generic MHD {\sc BATS-R-US} model \citep{powell99}. The model is driven by surface
magnetic field maps that are used to both determine the initial (potential) magnetic field
distribution as well as to scale the boundary conditions on the stellar photosphere.
In this model, a stellar wind solution is generated to be consistent with the distribution of the 
surface magnetic field.

Energy deposition in our simulations is assumed to be controlled by the expansion factor, $f_s$, 
of flux tubes, as in the case of the Sun. The basic assumption
is that the energization of each parcel of the stellar wind depends on the amount of expansion
of the flux tube from which this parcel comes from. \cite{wangy90} defined an
$f_s$ as the ratio between the magnetic flux at the source surface
(where the field becomes radial) and the magnetic flux at the photosphere for a particular flux tube.
Based on this definition and by fitting solar wind data, they derived an empirical relation
for $u_{sw} \propto \frac{1}{f_s}$, where $u_{sw}$ is the terminal wind velocity, $u(r\rightarrow \infty)$, for each
flux tube. This inverse relation states that slow wind comes from regions of large expansion of the flux tube
(the boundary between open and closed field lines),
while fast wind comes from regions with small expansion (nearly parallel open field lines) .
An improved empirical formula has been presented by \cite{argepizzo00} and is known
as the Wang-Sheeley-Arge model (WSA). This model has been used to predict the solar wind distribution
based on the solar surface magnetic field and its potential field extrapolation.
However, it does not provide any physical parameters other than the wind speed and the
polarity of the interplanetary magnetic field.

In order to obtain the complete set of physical parameters from our MHD model, we use the WSA model 
to determine how much excess energy to apply in each part of the 
simulation domain such that the input wind distribution is recovered in the MHD solution. 
Assuming the conservation of energy along a streamline (or a flux tube), we can equalize the total energy 
of the stellar wind on the stellar surface and at infinity. The total energy of the wind far from the star 
is equal to the bulk kinetic energy of the plasma, while the total energy on the stellar surface equals to the 
enthalpy minus the gravitational potential energy:
\begin{equation}
\frac{u^2_{sw}}{2}=\frac{\gamma_0}{\gamma_0-1}\frac{k_bT_0}{m_p}-\frac{GM_\star}{R_\star},
\label{BI}
\end{equation}
where $\gamma_0$ is the surface value of the polytropic index, $\gamma$, $k_b$ is the Boltzmann constant, 
$m_p$ is the proton mass, $G$ is the gravitational constant, and $R_\star$ and $M_\star$ are the stellar 
radius and mass, respectively. $T_0$ is the surface temperature; this boundary condition is a free parameter 
in the model. In principle, $T_0$ is taken to be equal to the average temperature of the 
stellar corona, but we emphasis that this is only a parameter in the model, which does not attempt to 
describe the actual coronal temperature. The kinetic energy due to stellar rotation is omitted assuming 
that this integral describes the radial acceleration of the wind and that the azimuthal component does not 
contribute to this acceleration. The Bernoulli equation equalizes the two ends of the streamline and 
in principle, the azimuthal component of the kinetic energy could be estimated and be added to the equation. 
However, due to the lack of stellar wind observations to constrain our model, we choose to adopt the original, 
solar empirical input. We expect the dominant effect of the fast rotation to appear in the solution mostly due 
to the rotational forces applied on the coronal plasma. The energy equation in the MHD solution does 
include all components of the velocity.  

Eq.~\ref{BI} enables us to relate the value of $\gamma_0$ to 
the terminal speed, $u_{sw}$, originating from this point and is known from the WSA model. Observations 
reveal that the value of $\gamma$ close to the Sun is found to be close to unity, while it is close to 
$3/2$ in the solar wind \citep{totten95,totten96}. This is due to the fact that close to the Sun the plasma 
contains some amount of ``turbulent'' internal energy, which is released in the process of the wind acceleration, 
so that far from the Sun the plasma is much less turbulent with $\gamma$ close to $3/2$. The complete 
description of this concept can be found in \cite{roussev03b}. Based on this concept, and the relation 
$\gamma_0(u_{sw})$, we can determine a volumetric heating function $E_\gamma(\mathbf{r},\gamma_0)$, 
which deposits energy in flux tubes such that smaller values of $\gamma_0$ (which have larger ``internal'' energy),
lead to a faster wind speed.

Once $E_\gamma$ is specified, the model solves the set of conservation laws for mass, momentum, magnetic induction,
and energy:
\begin{eqnarray}
&\frac{\partial \rho}{\partial t}+\nabla\cdot(\rho \mathbf{u})=0,&  \nonumber \\
&\rho \frac{\partial \mathbf{u}}{\partial t}+
\nabla\cdot\left(
\rho\mathbf{u}\mathbf{u}+pI+\frac{B^2}{2\mu_0}I-\frac{\mathbf{B}\mathbf{B}}{\mu_0}
\right) = \rho\mathbf{g},& \nonumber \\
&\frac{\partial \mathbf{B}}{\partial t}+
\nabla\cdot(\mathbf{u}\mathbf{B}-\mathbf{B}\mathbf{u})=0, &  \\
&\frac{\partial }{\partial t}\left(
\frac{1}{2}\rho u^2+\frac{1}{\Gamma-1}p+\frac{B^2}{2\mu_0}
 \right)+ &
 \nonumber \\
&
\nabla\cdot\left(
\frac{1}{2}\rho u^2\mathbf{u}+\frac{\Gamma}{\Gamma-1}p\mathbf{u}+
\frac{(\mathbf{B}\cdot\mathbf{B})\mathbf{u}-\mathbf{B}(\mathbf{B}\cdot\mathbf{u})}{\mu_0}
\right)=\rho(\mathbf{g}\cdot\mathbf{u})+E_\gamma, \nonumber &
\label{MHD}
\end{eqnarray}
with $\Gamma=1.5$ (note that the parameter $\gamma$ that defines the empirical energy source term $E_\gamma$ is 
always $\le \Gamma$) until convergence is achieved. As in any other numerical
MHD model, the condition of $\nabla\cdot\mathbf{B}=0$ needs to be enforced throughout the simulation.
In this model, the ``eight wave'' method is used for divB cleaning \citep{powell99,toth00}.

The necessary inputs for the model are the surface distribution of the radial magnetic
field, the boundary value for the density, $\rho_0$, and the temperature, $T_0$,
as well as the stellar radius, $R_\star$, mass, $M_\star$, and rotation period, $\Omega_\star$.
The stellar properties used in the simulation are presented in Table~\ref{table:t1} \citep[see the summary of][]{Drake08}.
The two boundary conditions ($\rho_0$ and $T_0$) are chosen so that they are slightly higher than the
typical values used for the solar case, as would be expected for an active star. 
One can argue that the empirical wind speed calculated in the WSA model is fitted to the 
solar case, and that it should not be used for stellar applications. However, due to 
the lack of stellar wind observations, we choose to use of the solar wind input 
over the use of a few indirect observation and general scaling laws for stellar winds \citep{wood04,Wood05}, 
which hold many other uncertainties. In any case, the range of magnitude and distribution of the wind 
from the empirical formula in the WSA model can be easily modified, but this parameterization is out of 
the scope of this paper. In addition, the basic assumption here is that as long as the process of wind 
acceleration is assumed to be similar to the Sun, the relation $u_{sw}(1/f_s)$ holds. 
Since a complete theoretical model for wind acceleration is unavailable even for the Sun, we argue 
that our approach can provide an insight about the structure of stellar coronae, despite of the many assumptions 
it holds.

\subsection{SIMULATION SETUP}
\label{sec:Model2}

\begin{figure*}[h!]
\centering
\includegraphics[width=6.in]{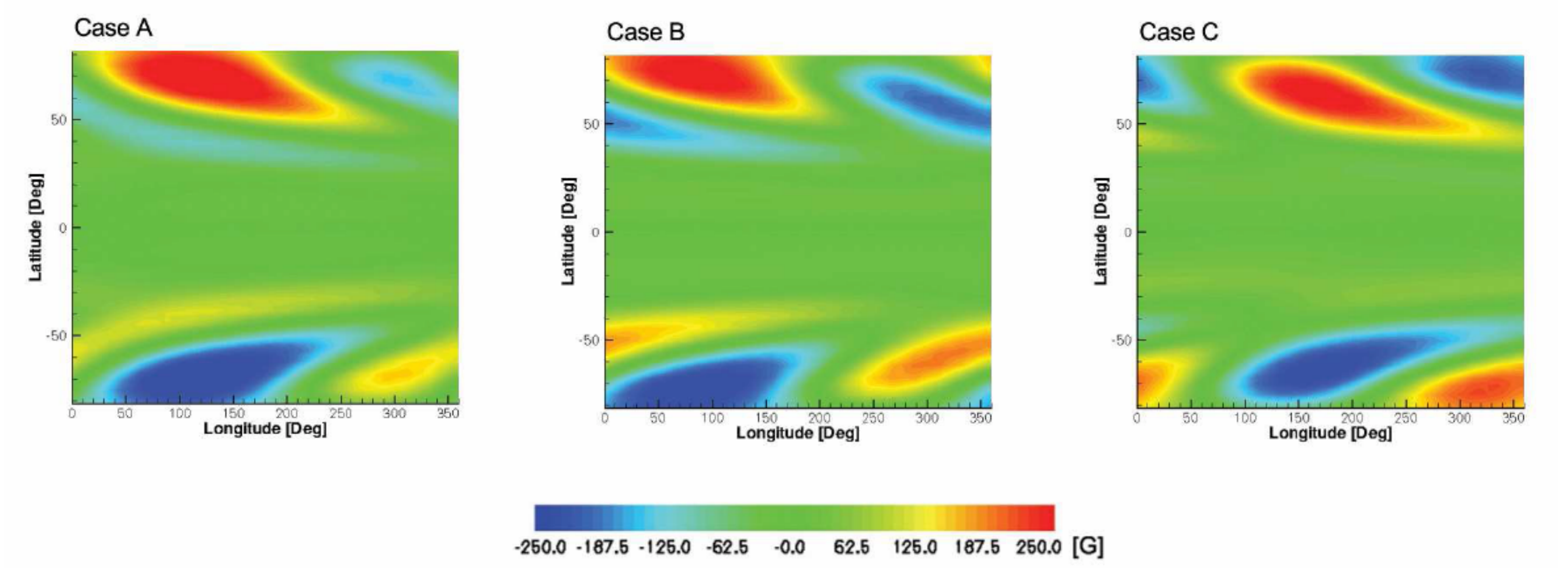}
\caption{Input magnetic surface maps used in Cases A, B, and C (left to right) obtained from the stellar
``flip-flop" dynamo model for FK~Com \citep{Elstner05,KorhonenElstner05}.
These cases cover three different regimes of the dynamo cycle: dominance of each hemisphere by a single spot (A); two spots in each hemisphere of equal magnetic field strength (C); and the intermediate case in which there are two significant spots but with one dominant (B).
}
\label{fig:f1}
\end{figure*}

FK~Com rotates so rapidly that its absorption lines are strongly smeared, presenting a severe challenge to 
Zeeman-Doppler imaging techniques that have been successfully applied to other stars \citep[e.g][]{Donati09,Donati09a}.
Consequently, surface magnetic field maps are not available and we use the surface field
predicted by \cite{Elstner05} and \cite{KorhonenElstner05} based on a dynamo simulation of the
system, where the maximum value of $\approx 250\;G$ is in agreement with recent magnetic field observations 
of the system \citep{Korhonen09}. The dynamo model predicts the temporal evolution of the stellar field, and for the
purpose of this paper, it provides the surface distribution for different phases
of the flip-flop cycle.  We simulate the coronal structure of the FK~Com model for
three different phases, at which there are distinct differences in surface magnetic topology.  
Figure~\ref{fig:f1} shows the magnetic maps used for the three cases, which we call ``Case A'', ``Case B'', 
and ``Case C''. Case A corresponds to the field being concentrated in two single large spots of opposite polarity 
located in opposite latitudinal hemispheres. Case C has a more equal division of field in two spots of opposite polarity 
in each latitudinal hemisphere, again mirrored in polarity about the equator (so the total number of spots is 4). Case B is intermediate between the two, in 
which there is one dominant spot in each latitudinal hemisphere, but with a second spot containing significant field of opposite polarity. 
We do not consider the cases differing to these in phase by 180 degrees in the magnetic cycle,  since these would simply repeat the magnetic structure of cases 
A-C mirrored about the equator.  Our main focus is to study the global effect of fast rotation on the coronal structure of these models.
The maps used here have low resolution relative to solar maps, so we do not
need to capture small active regions like in the solar case, and the
simulation described here does not require very high resolution in order to
capture the global coronal structure.  The smallest grid size
near the surface of the star is $\Delta x=3\cdot10^{-2}R_\star \approx 2\times10^{10}$~cm, and the total
number of cells used is $3\cdot 10^6$.

We ran the simulation in the frame of reference rotating with the
star using the local time step algorithm \citep{cohen08b}. This enables a much
faster convergence for steady state simulations. The limitation of this approach
is that the rotational forces become too large at large distances when simulating
a very rapidly rotating system, and the simulation becomes numerically unstable. Here,
since we are interested in the low coronal structure, we limit the
Cartesian simulation domain to extend only up to $15R_\star$ in each direction.


\section{RESULTS}
\label{sec:Results}

\begin{figure*}[h!]
\centering
\includegraphics[width=6.in]{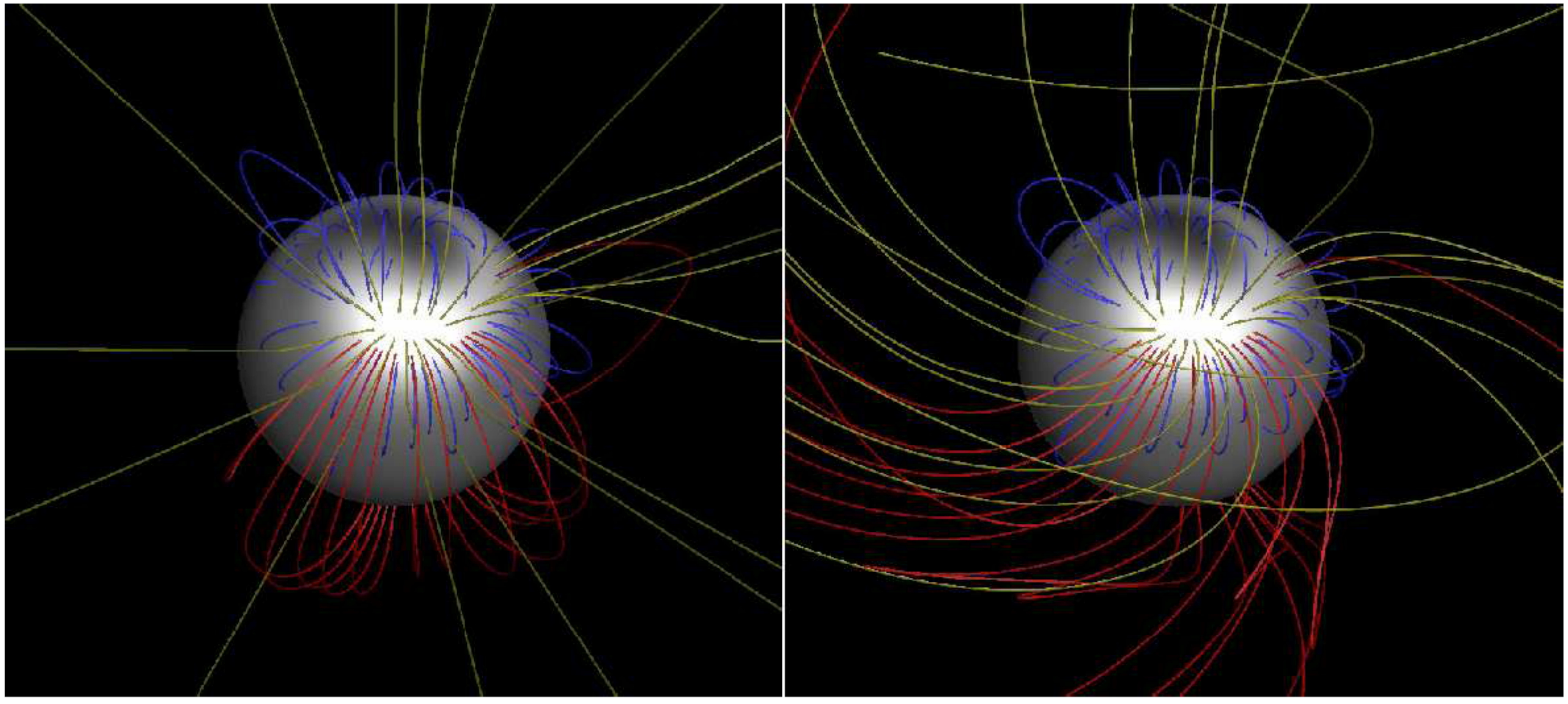}
\caption{Potential field extrapolation (left), and non-potential MHD solutions (right) for Case A. Blue field lines are
closed in both solutions, yellow field lines are open in both solutions, and red field lines are
field lines that their topology drastically changes from the potential field to the MHD solution.
The stellar surface is colored according to the magnitude of the
perpendicular surface magnetic field.}
\label{fig:f2}
\end{figure*}

We first compare a potential
field extrapolation with the non-potential fields from the corresponding stationary MHD simulation in Figure~\ref{fig:f2}.
The potential field extrapolation (left panel) has a source surface (not shown in the zoomed in figure) located at
$2.5R_\star$, and a number of field lines, both open and closed, are shown.
Field lines that cross the source surface are
fully open, as required by the boundary condition mentioned in \S\ref{sec:Intro}.
Similarly, field lines corresponding to the same footpoint locations are also shown
for the non-potential field from the MHD solution for Case A (right panel; the other
cases show qualitatively similar behavior, as we show below in Figure~\ref{fig:f3}).  The field
lines are color-coded based on their behavior in the two solutions: those that are
closed in both are colored blue, those that are open in both are in yellow, and
those that change their topology between the potential and MHD solution are shown in red. 
The blue, closed field lines are all quite low-lying in both solutions, though are significantly 
more stretched out in the MHD solution.  The behavior of the red field lines is instead quite striking: 
these represent a class of field lines that are stretched and twisted when a wind flow and rotation are 
imposed on the system.

Another important difference between the potential field and MHD solutions is the
effect of the fast stellar rotation on the geometry of the field lines.  
Due to the (infinite) conductivity of the plasma in the MHD solution, the 
magnetic field lines are frozen in to the plasma, which propagates radially in the inertial frame.
The footpoints of the field lines 
are attached to the rotating stellar surface. As a result, the field is wound up and stretched 
around to form an enhanced, compact version of the Parker spiral \citep{Parker58}. A major difference here 
is that in the solar case, the toroidal component of the coronal field becomes dominant 
{\em beyond} the Alfv\'en point.  In the case of rapidly rotating stars, as exemplified by our simulations, the toroidal component 
is strong {\em inside} the Alfv\'en point, and can feedback on the system.   This strong 
toroidal component cannot be predicted by the potential field approximation.

The global structure of the magnetic field and the radial velocity fields derived
from the MHD simulations are shown in Figure~\ref{fig:f3} for Cases A (top),
B (middle), and C (bottom).  The panels show slices in the $u_r$ field at
different orientations: one to indicate the 3D structure by showing the
solutions in the $y=0$ and $z=0$ planes (left), and one to show
the structure in the $y=0$ plane containing the rotation axis (right).
The stellar surface is shown colored with the input radial magnetic
field, as in Figure~\ref{fig:f2}, and the 3D magnetic field lines are
shown in white.  The differences in the structure of the velocity fields
in the three cases are more apparent in the right panel figures, where there is 
a clear correlation between the distribution of the radial stellar wind and the 
surface magnetic topology. In Case A, 
we have a single dominant fast stream in each latitudinal hemisphere associated with the main strong spots. In case 
C, we obtain three fast streams together with a weaker fast stream, all associated with the 
equally strong surface spots. In case B, the magnitude of the radial flow decreases, since in this case the 
surface spots are weaker than cases A and C.  The weak outflow component of 
the flow introduces a more turbulent solution and a strong inflow component in the southern hemisphere. 
Another notable feature is the fact that magnetic field lines that originate from regions close 
to the strong spots have a shape similar to the one predicted by the Parker spiral (due to the 
strong outflow component), while field 
lines that originate from regions far from the strong spots (where the outflow component is weaker) 
are more likely to be affected by local changes of the flow. 

The stellar wind structure is determined by the topology of the open/closed field lines and the expansion of the flux tubes. 
In the solar case, where the solar rotation is not extreme, the input speed from the WSA model is well reproduced by 
the MHD solution \citep{cohen08b}. In the case of FKCom however, we
use the empirical model to specify the acceleration of the radial wind, while we use the MHD 
model to study the effect of fast rotation on the coronal structure. This part cannot be captured by 
the empirical model. The difference in the stream structures for the three cases can be
attributed to the different density and magnetic field  structures present in the stationary
solutions.  These are illustrated in Figure~\ref{fig:f4} that shows the number density close to the star
in the $y=0$ plane, as well as an isosurface of constant density,
$n=1\cdot10^9$~cm$^{-3}$ (green surface). It can be seen that the large streamers in Case B 
(middle panel) are more stretched than case A (left panel) and case C (right panel). 
This reduces the number of flux tubes associated with open field lines (with small expansion) and, as a result, 
the radial outflow component decreases as seen in Figure~\ref{fig:f3}.


\section{DISCUSSION}
\label{sec:Discussion}

\begin{figure*}[h!]
\centering
\includegraphics[width=7.in]{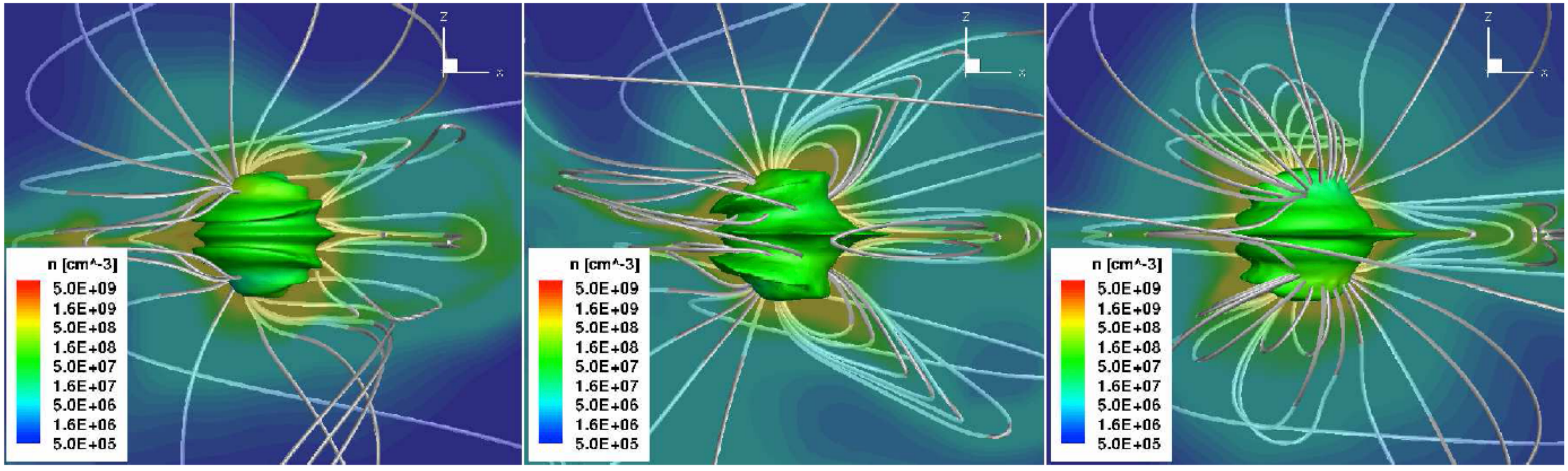}
\caption{Density structure and configuration of the steady state solutions for Cases A (left), B (middle), and C (right) close to the star. Color contours are
of number density, and the green surface represents an iso surface of $n=1\cdot 10^9~cm^{-3}$. The 3-dimensional
magnetic field lines are shown in white.}
\label{fig:f4}
\end{figure*}

Based on the simulation results, we emphasize two aspects of the results.
First, the non-potential, MHD steady state solution is significantly different from the
potential field extrapolation for the FK~Com-like flip-flop dynamo system.  Second, the coronal
structure of such an FK~Com-like star is expected to be qualitatively different from the solar corona
and coronae of more slowly rotating Sun-like stars.  The former is evident from the class of field lines
that change their topology dramatically from the potential field to the MHD
simulation (Figure~\ref{fig:f1}), and the latter can be seen in the large-scale 
topology of the coronal magnetic field (Figure~\ref{fig:f3}-\ref{fig:f4}).
This structure is dominated by the tangling of the rotationally-wound, large-scale magnetic field, unlike the
solar case which is dominated by coronal holes and active regions.

The rapidly rotating plasma environment acts to inhibit the radial
component of the flow, and the density decrease with radial distance is less
pronounced than would be expected in Sun-like coronae. The strong rotational
component causes a toroidal ``dragging" and stretching of the field
lines, which nevertheless remain closed since the azimuthal forces do not
overcome the magnetic tension as in the radial case.  This stretching, however, stores
magnetic energy in the loops and causes an increase in the magnetic
tension, $T_B=\mathbf{B}\cdot\nabla\mathbf{B}/\mu_0$, over the entire length of 
the loops. The magnetic tension is one contributer to the Lorentz force (the other 
one is the magnetic pressure) and it has the SI units of $Pa\;m^{-1}$ (momentum). 

\begin{figure*}[h!]
\centering
\includegraphics[width=6.in]{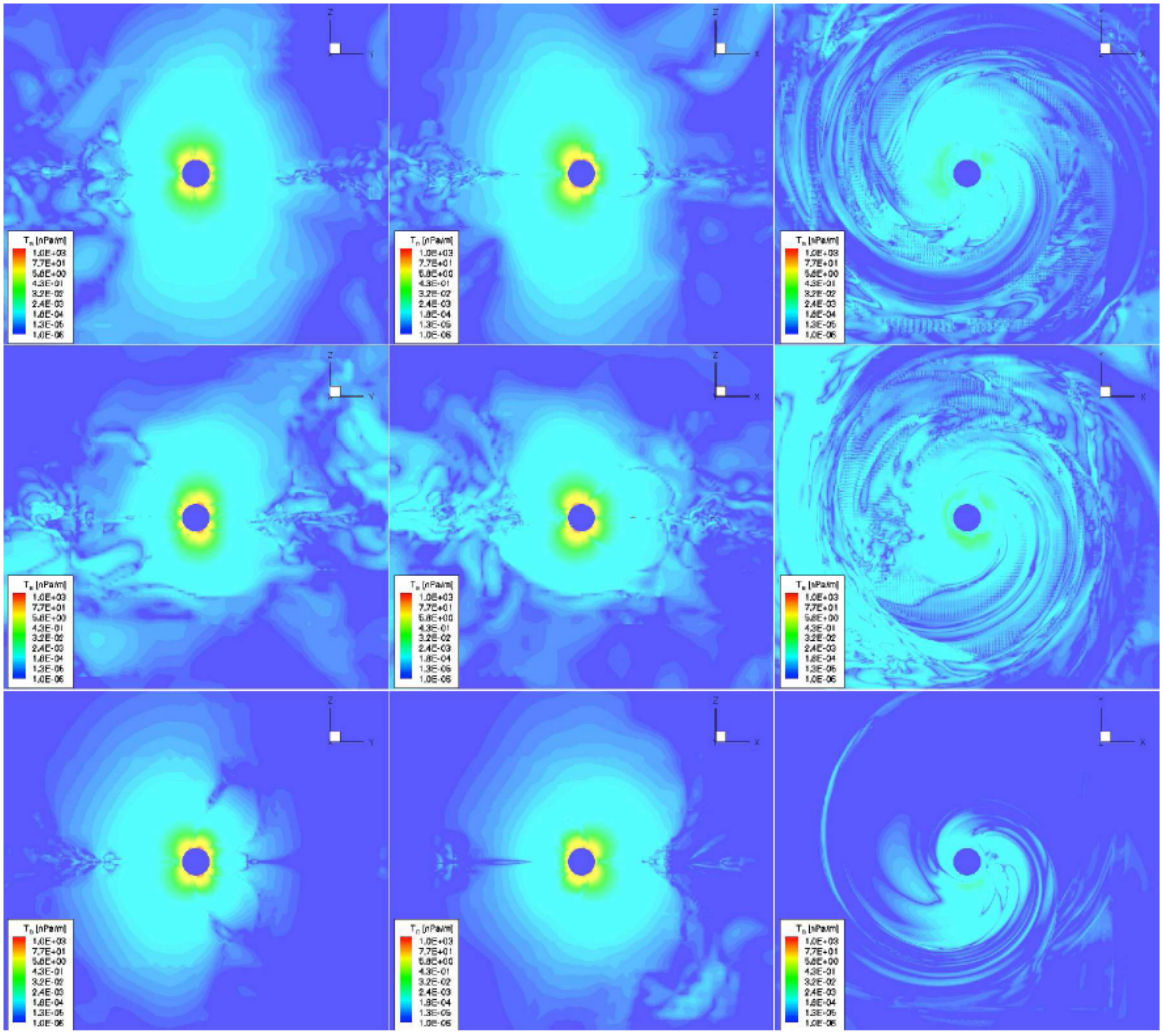}
\caption{Magnetic tension on the $x=0$ (left), $y=0$ (middle), and $z=0$ (right) for cases A-C (top to bottom). }
\label{fig:f5}
\end{figure*}

Figure~\ref{fig:f5} shows maps of the magnetic tension in $nPa\;m^{-1}$ on the $x=0$ (left), $y=0$ (middle), 
and $z=0$ (right) planes for cases A-C (top to bottom). The magnetic tension in the maps is controlled by 
the interplay of two components. The outflow component that stretches the field lines radially, and the azimuthal 
component of the flow that stretches the closed loops in the azimuthal direction. In case A, we have 
two main fast radial streams and the signature for them can be seen closer to the star, especially in the 
$y=0$ plane.  Case B is characterized by a weaker outflow component, so that the azimuthal component takes over to 
dominate the distribution of the magnetic tension. In particular, the magnetic tension on the $z=0$ plane in 
this case is stronger, indicating a stronger stretch applied on the equatorial loops.  In addition, the magnetic 
tension is weaker in the southern hemisphere due to the lack of strong outflow in these regions. Case C is characterized 
by four outflows at higher latitudes, so the magnetic tension near the equator is weaker than the other two cases.

\begin{figure*}[h!]
\centering
\includegraphics[width=6.in]{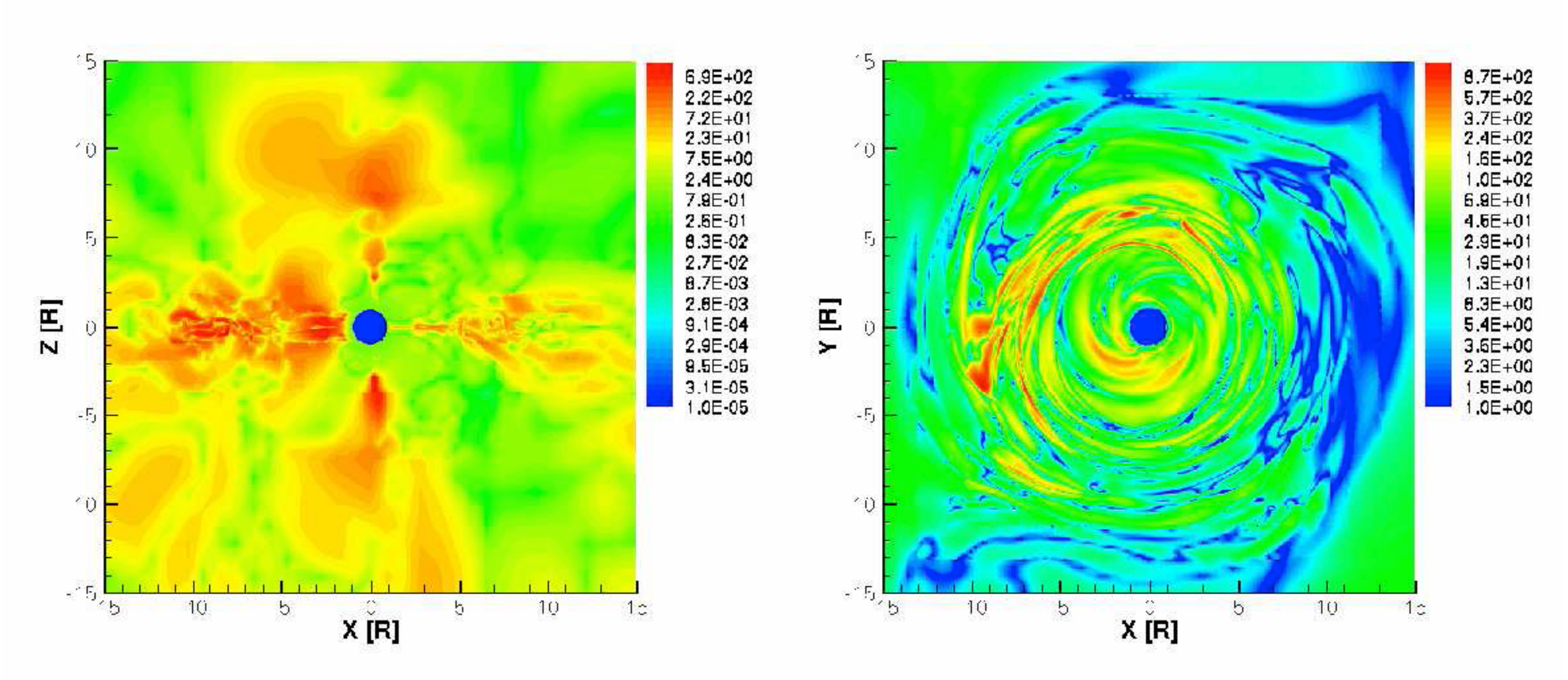}
\caption{Ratio of $T_B$ in the MHD solution over $T_B$ in the potential field displayed on the $y=0$ (left) and $z=0$ (right) plains for Case C.}
\label{fig:f6}
\end{figure*}

In Figure~\ref{fig:f6}, we show the distribution of the ratio of $T_B$ in the MHD solution to that in the 
potential field, for Case C on the $y=0$ and $z=0$ planes (the overall behavior for Cases  A and B is similar).  
The signature of the stretched field lines is easily visible in the equatorial plane ($z=0$) plots. 
The magnetic tension in the $y=0$ plane is mostly due to the 
radial gradient in $\mathbf{B}$ and, therefore, it is quite similar in the potential and non-potential solutions. 
The magnetic tension is highest near the surface, particularly near regions of spot activity 
where active regions are expected to exist.  These are regions where the
loop footpoints will be subjected to considerable stress due to convective
motions in the photosphere \citep{Leighton64,Parker74,Fisk05}.  Such sites are conducive
to enhanced activity and the generation of X-ray flares.
In the simulation presented here, the corona reaches a steady state so that
the stretched loops remain unchanged. It is likely, however, that time-dependent
processes observed on the Sun, such as footpoint surface motions and evolving
small active regions exist on FK~Com-type stars that can trigger stellar eruptions and flares.
It is also possible that these processes, in particular loop footpoint motions,
can lead to magnetic reconnections between the stretched loops themselves
and to major flaring activity on a global scale from the star.  The study of these processes
would be of great interest for rapidly rotating coronae but cannot be addressed
in the stationary simulations presented here.

We expect that the strong azimuthal wrapping of the coronal magnetic field, {\em within the Alfv\`en radius}, 
will be a general feature of the coronae of rapidly rotating stars of all spectral types.  We also draw attention 
to the closed, wrapped magnetic field structures evident in the left panels of Figure~\ref{fig:f3} and in 
Figure~\ref{fig:f4}. The apex of such a field line lies between one to several stellar radii from the stellar surface. 
This type of structure would be a likely candidate for hosting the ``slingshot'' prominences commonly found on rapidly 
rotating dwarfs \citep[e.g][]{Robinson86,Collier-Cameron89a,Collier-Cameron89b,Collier-Cameron92}.


\section{SUMMARY AND CONCLUSIONS}
\label{sec:Conclusions}

We have performed global MHD simulations for an FK\,Com-like system 
in order to study the
three-dimensional structure of the corona of very rapidly rotating stars.  We drove the model using surface
synthetic magnetic maps produced by a stellar ``flip-flop" dynamo simulation of FK~Com.

Our main findings are:
\begin{enumerate}
\item For rapidly rotating systems, the potential field extrapolation is
inadequate to describe the structure of the coronal magnetic fields;
\item The simulations show the presence of a significant azimuthal component in
the coronal flow that impedes the radial component (the stellar wind outflow);  
\item The fast stellar rotation combined with the highly conductive outflowing wind generates a 
strong toroidal component of the magnetic field in the form of highly stretched coronal loops within the Alfv\'en radius and wrapped open field lines beyond;
\item This stretching introduces regions of large magnetic tension in the stationary solutions which may act as
sites where magnetic reconnection can be preferentially triggered via surface
footpoint motions and emergence of new flux. A disconnection of such stretched loops might lead to major, large-scale stellar flares.
\end{enumerate}

The simulation presented here provides a first glimpse of the structure
of the corona of a rapidly rotating star. We expect the salient features of our results to be generally 
applicable to rapidly rotating dwarfs as well as FK~Com-type stars. Of interest for future study 
would be the distribution and magnitude of the predicted wind mass flux, and the influence on this of 
the changing coronal topology through the dynamo cycle.  
In addition, time-dependent study of how the azimuthally stretched loops interact and influence magnetic 
reconnection and the evolution of coronal mass ejections would be highly motivated.


\acknowledgments

OC is supported by SHINE through NSF ATM-0823592 grant, and by NASA-LWSTRT Grant NNG05GM44G.
JJD and VLK were funded by NASA contract NAS8-39073 to the {\it Chandra X-ray Center}.
Simulation results were obtained using the Space Weather Modeling
Framework, developed by the Center for Space Environment Modeling, at the University of Michigan with funding
support from NASA ESS, NASA ESTO-CT, NSF KDI, and DoD MURI. 




\begin{table}[h!]
\caption{Adopted Properties of FK~Com .}
\begin{tabular}{c||c}
\hline
$\rho_0$ & $5\cdot 10^{9}\;cm^{-3}$\\
$T_0$ & $7\;MK$\\
$R_\star$ & $8.7\;R_\odot$ \\
$M_\star$ & $2.2\;M_\odot$ \\
$P_{rot}$ & $2.4\;d$  \\
\hline
\end{tabular}
\label{table:t1}
\end{table}
\clearpage


\clearpage

\begin{figure*}[h!]
\centering
\includegraphics[width=6.0in]{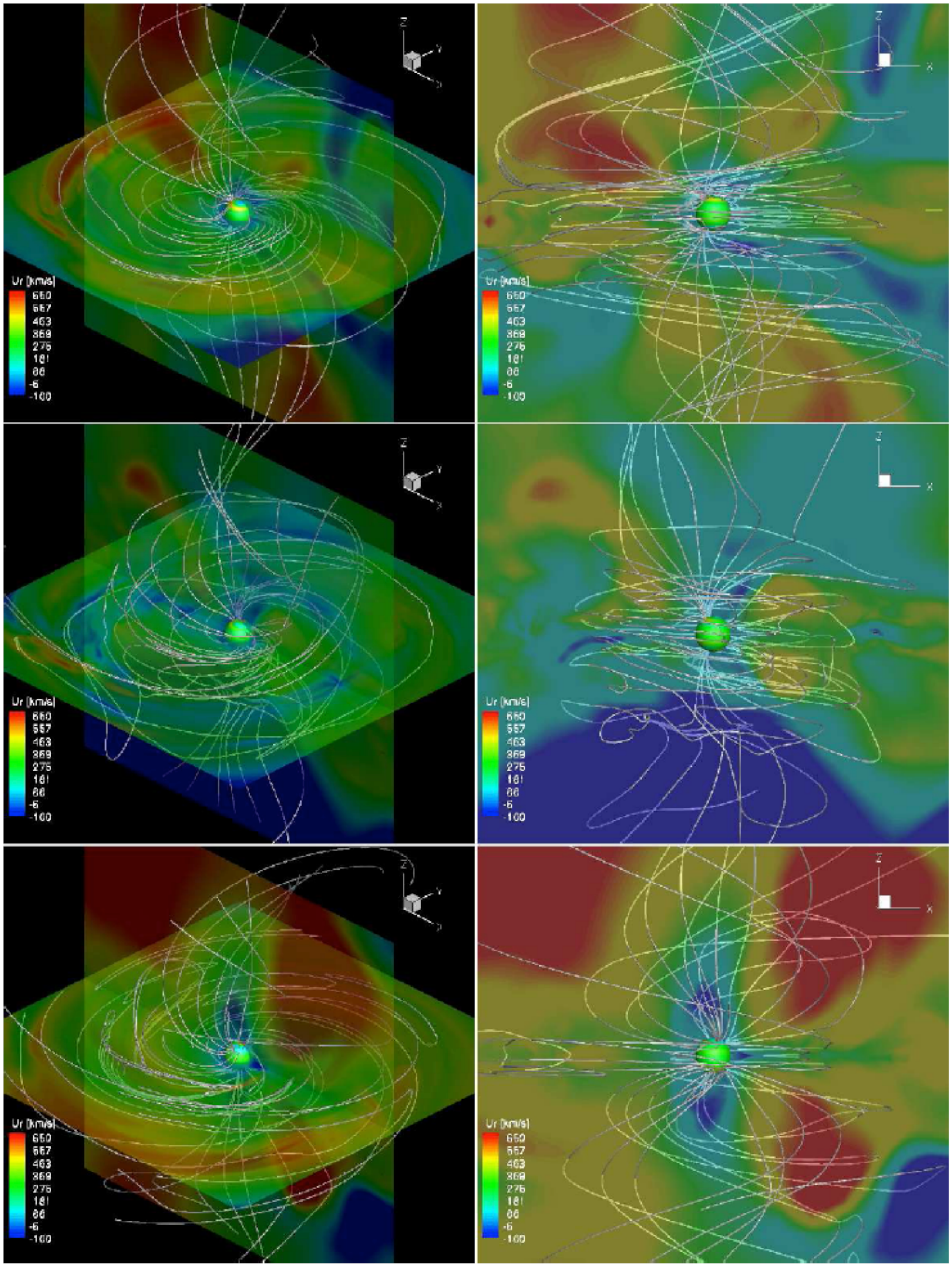}
\caption{Global views of the stationary magnetic field solutions for the
three cases A (top), B (middle), and C (bottom).
The 3-dimensional magnetic field lines are shown in white.
Color contours of $u_r$ are displayed on the $y=0$ and $z=0$ planes (left panels).
The right panels show a similar display, but with a side view of the $y=0$ plane
for clarity.
The stellar surface is also shown, colored according to the magnitude
of the surface field as in Figure~\ref{fig:f2}.}
\label{fig:f3}
\end{figure*}
\clearpage

\clearpage

\end{document}